\documentclass{iopconfser}
\usepackage{graphicx}
\usepackage{subcaption}
\usepackage{hyperref}
\hypersetup{
    colorlinks=true,
    linkcolor=blue,
    filecolor=magenta,      
    urlcolor=cyan,
    pdftitle={JUNO DC-1},
    pdfpagemode=FullScreen,
    }

\usepackage{lineno}
\modulolinenumbers[1]
    
\begin{document}

\title{Offline data processing in the First JUNO Data Challenge}



\newcommand{\IHEP}{1}
\newcommand{\UCAS}{2}
\newcommand{\JINR}{3}
\newcommand{\SYSU}{4}
\newcommand{\SDU}{5}
\newcommand{\IPHC}{6}
\newcommand{\inst}[1]{$^{#1}$}

\author{Tao Lin\inst{\IHEP}, 
    Weiqing Yin\inst{\IHEP, \UCAS} 
    (on the behalf of JUNO Collaboration)}

\affil{\inst{\IHEP}Institute of High Energy Physics, Beijing, China}
\affil{\inst{\UCAS}University of Chinese Academy of Sciences, Beijing, China}
\email{lintao@ihep.ac.cn}

\begin{abstract}
The Jiangmen Underground Neutrino Observatory (JUNO) is currently under construction and the installation of detector will be completed by end of 2024. A series of JUNO Data Challenges are proposed to evaluate and validate the complete data processing chain in advance. In this contribution, the offline data processing in the first JUNO Data Challenge (DC-1) is presented. The primary goal of DC-1 is to process one week data using conditions database and multi-threaded reconstruction. The workflow involves the production of simulated data and reconstruction of the data. To achieve the goals, a JUNO-Hackathon has been organized. The software performance is measured and the results are presented.
\end{abstract}

\section{Introduction}

The Jiangmen Underground Neutrino Observatory (JUNO) is a multipurpose neutrino experiment with the primary goals of the determining the neutrino mass ordering and precisely measuring oscillation parameters \cite{An:2015jdp,Djurcic:2015vqa,JUNO:2021vlw}. Currently under construction in Southern China, it comprises a central detector (CD) for neutrino detection, a water pool (WP) and a top tracker (TT) for cosmic ray muon measurement. The innermost of CD is 20 kton liquid scintillator (LS), surrounded by 17,612 20-inch and 25,600 3-inch photomultiplier tubes (PMTs). The WP is equipped with 2,400 20-inch PMTs, served as a veto system for the cosmic ray muons. On the top of WP, the TT is used to measure the muons as well. The installation of all the PMTs and readout electronics will be completed by the end of 2024 and that afterwards, during detector filling, JUNO will start commissioning all the readout channels to test the full DAQ and data processing chains.

\begin{figure}[h]
    \centering
    \includegraphics[width=0.6\linewidth]{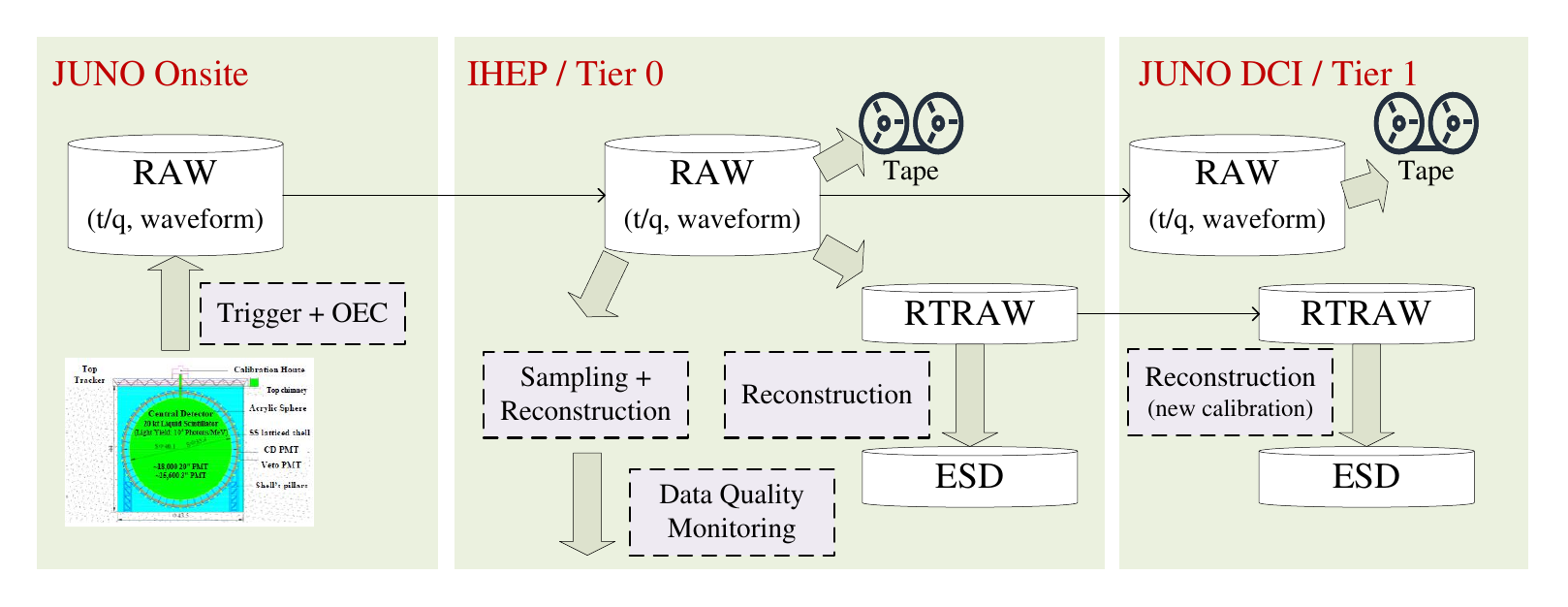}
    \caption{Baseline scheme of offline data processing \label{fig:scheme}}
\end{figure}

Figure \ref{fig:scheme} shows a schematic view of data processing in JUNO. When data taking starts, the event rate is about 1 kHz. The detector produces thousands channels of waveforms at a sampling rate of 1 GHz. To reduce the huge data volume, an additional system named Online Event Classification (OEC) is applied after trigger to reduce the event size according to the event types. Unlike the trigger system discards events, OEC retains all events. Approximately 60 MB/s of byte-stream RAW data, amounting 2 PB per year, is expected to be produced. The RAW data is transferred from the onsite to the IHEP data center via a dedicated network. The RAW data is preprocessed and converted to the ROOT-based RAW (RTRAW) data, using JUNO Event Data Model \cite{Li:2017zku} and ROOT I/O\cite{Brun:1997pa}. Both types of data are replicated to the other data centers through the Distributed Computing Infrastructure (DCI) \cite{Zhang:2024anf}. To minimize disk volume, the RAW data is archived to a tape library, while the RTRAW data is stored on a disk. After event reconstruction of the RTRAW data, the output is stored in ESD (Event Summary Data) format. The data processing involves several critical components, including data quality monitoring (DQM), keep-up reconstruction (KUP) and physics production (PP). 

\section{The first JUNO Data Challenge (DC-1)}

A series of JUNO Data Challenges (DC) have been proposed to evaluate and validate the complete data processing chain in advance. The JUNO DC serves not only to test the event reconstruction software, but also functions as a system test for the database, the Kafka-based \cite{10.5555/2588385} data pipeline, DQM, KUP and PP etc. The estimation of computing and storage capacities are validated through JUNO DC.

The JUNO DC-1 is focus on data processing within the central detector. About one week of inclusive datasets are produced and then reconstructed. Radioactivities, cosmic ray muons and neutrino events are simulated in advance. The rates of radioactivities and muons are set to the rates expected in the real data, while the rate of neutrino events is increased from 60 events per day to 4 Hz. By increasing the event rates of neutrinos, the reconstruction algorithms can be tested with higher statistics. In order to test the conditions database later, seven sets of time offsets are added to the channels during simulation. 

\begin{figure}[h]
    \centering
    \includegraphics[width=0.6\linewidth]{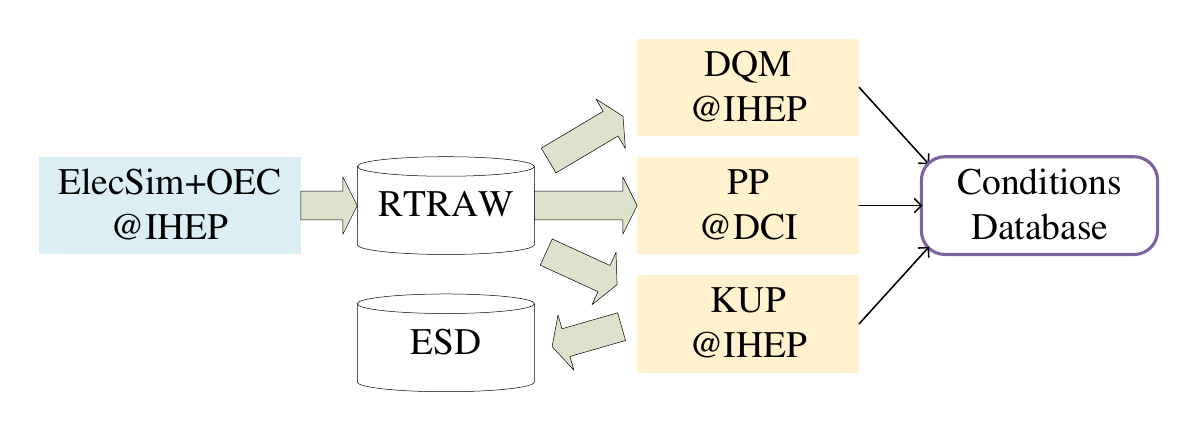}
    \caption{Workflow in DC-1 \label{fig:workflow}}
\end{figure}

As shown in Figure \ref{fig:workflow}, there are two major steps in the workflow. The first step involves using simulation software to generate RTRAW files. The subsequent step involves reconstruction of these RTRAW files in the different systems and producing ESD files. Additionally, the conditions database is used for testing. 

\begin{figure}[h]
    \centering
    \includegraphics[width=0.5\linewidth]{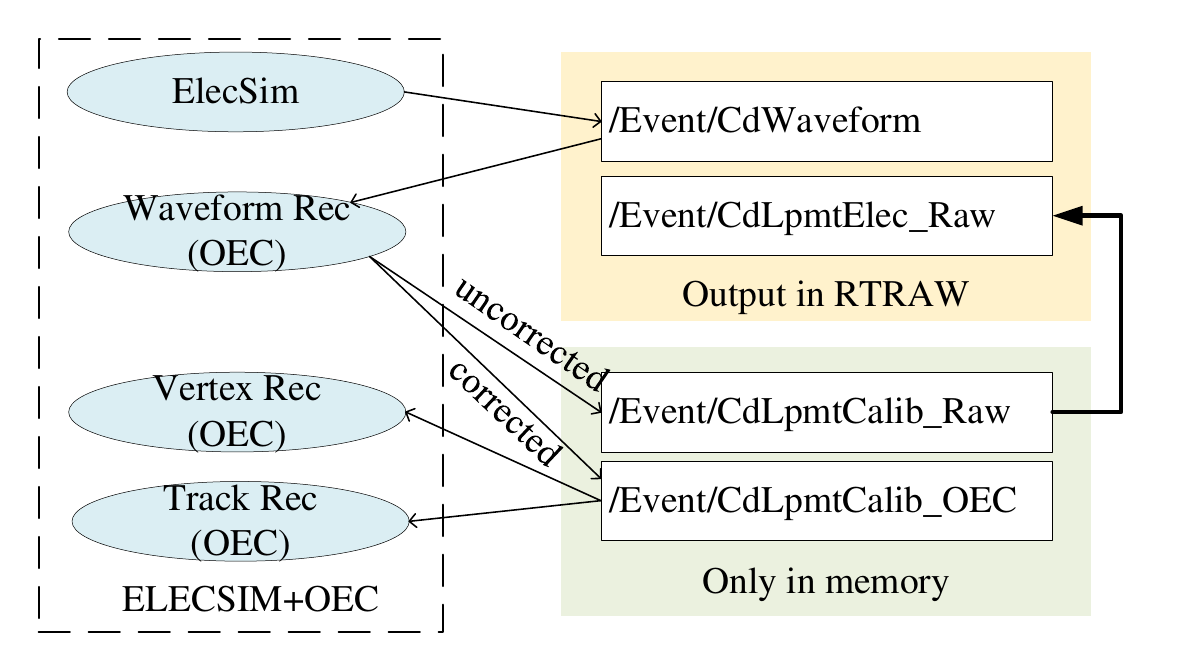}
    \caption{Workflow in ElecSim and OEC \label{fig:workflowRTRAW}}
\end{figure}

The JUNO simulation software \cite{Lin:2022htc,Lin:2017usg} incorporates Geant4-based \cite{GEANT4:2002zbu,Allison:2006ve,Allison:2016lfl} detector simulation and electronics simulation with OEC. As previously mentioned, the existing detector simulation datasets, which serve as inputs of the electronics simulation, are produced prior to the electronics simulation and are mixed according to the event rates. The electronics simulation generates pulses for all detector channels, which are subsequently digitized into waveforms after trigger. Unlike collider experiments, time correlation is curcial, which requires that the events could not be discarded. To reduce the data volume, OEC is employed to classify the event types using fast reconstruction results and to select a storage strategy for each event. Multiple events within a given time window are used for event type classification, a process known as time correlation analysis. As shown in Figure \ref{fig:workflowRTRAW}, the waveforms are reconstructed to time and charge (t/q) information in the OEC first. This t/q information is then calibrated, and the events are reconstructed with the calibrated t/q information. The OEC determines whether the waveform or t/q information should be stored in the final output file. For instance, waveforms are stored for neutrino events, while only the t/q information is stored for the other types. For the t/q stream, only the uncorrected data is stored into file, permitting offline correction with the conditions database at a later stage. 

\begin{figure}[h]
    \centering
    \begin{subfigure}{0.48\textwidth}
        \includegraphics[width=\textwidth]{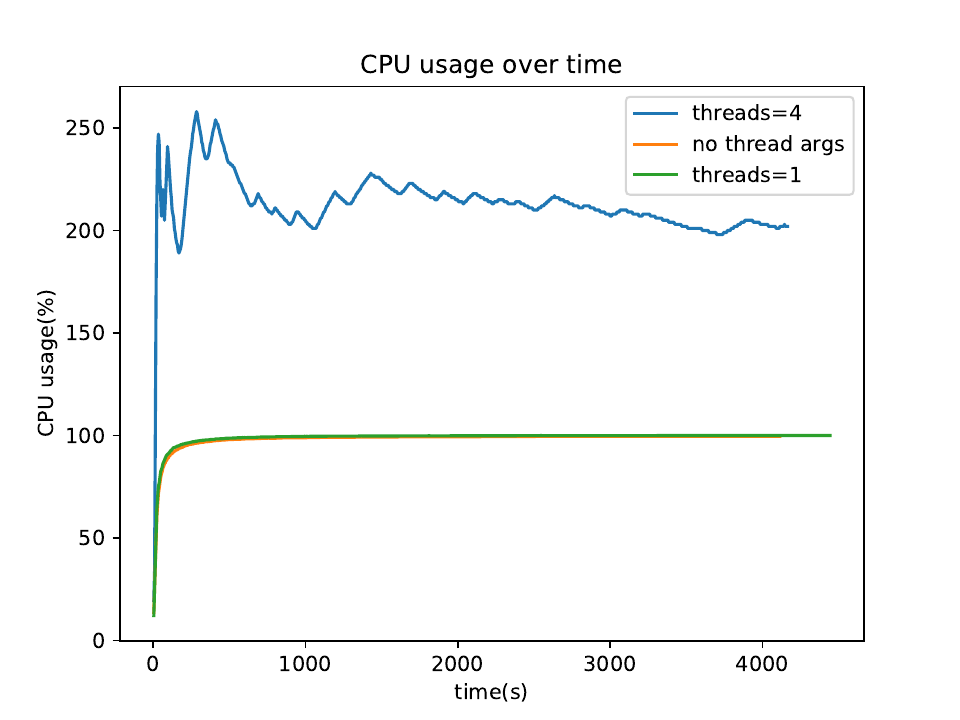}
    \end{subfigure}
    \begin{subfigure}{0.48\textwidth}
        \includegraphics[width=\textwidth]{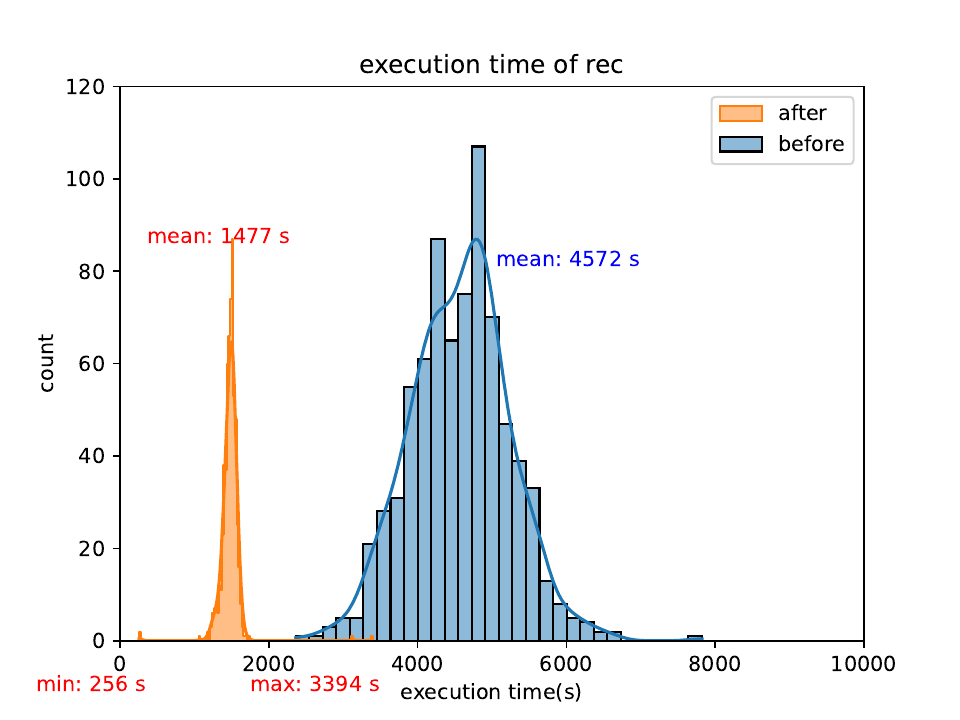}
    \end{subfigure}
    \caption{(a) Low CPU usage due to an issue in multi-threaded track reconstruction. (b) Execution time before and after issue fixed. \label{fig:perfMTRecLowCPU}}
\end{figure}

JUNO Hackathon was organized with the aim of migrating the reconstruction algorithms and conditions database from a serial version to a multi-threaded version. Any issues that encountered during testing have been addressed and resolved. Profiling tools, such as Intel VTune \cite{vtune}, were used to identify bottlenecks. Figure \ref{fig:perfMTRecLowCPU} illustrates one of the issues related to low CPU usage, as well as the CPU usage after optimization. A significant issue was the internal use of locks during the event processing, which resulted in multiple threads being blocked when attemping to access the locks. 

Dedicated computing resources are employed in JUNO DC-1. Since the data taking has not yet started, the dedicated DQM cluster is used for the testing purposes. This cluster comprises 36 computing nodes with a total of 2304 cores, which are managed by the HTCondor system \cite{HTCondor_Team_HTCondor}. A total of 576 job slots are allocated, with each slot equipped with 4 cores and 15 GB memory. 

\section{Software performance}

\subsection{RTRAW production}
\begin{figure}[h]
    \centering
    \begin{subfigure}{0.48\textwidth}
        \includegraphics[width=\textwidth]{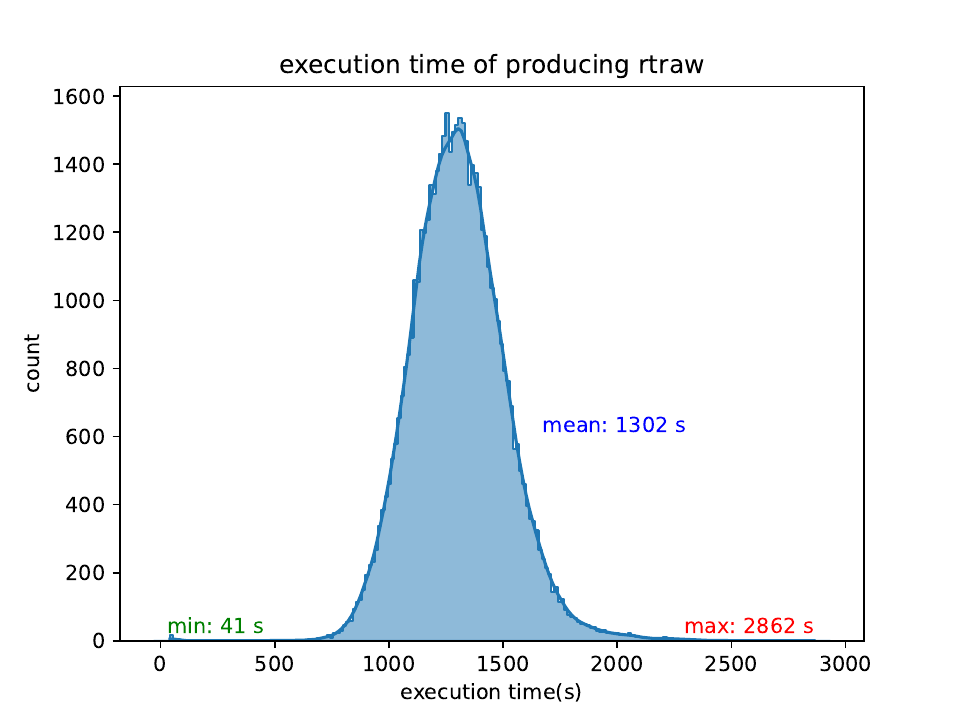}
    \end{subfigure}
    \begin{subfigure}{0.48\textwidth}
        \includegraphics[width=\textwidth]{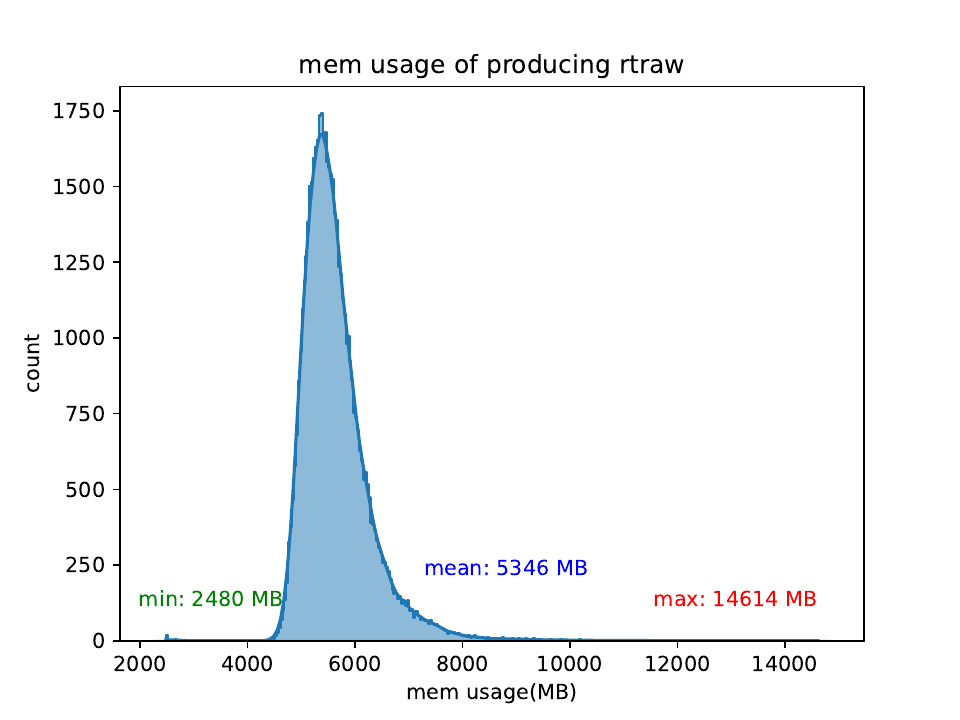}
    \end{subfigure}
    \caption{Performance of generating 6s RTRAW data. (a) Execution time of the simulation. (b) The maximum memory usage during simulation. \label{fig:perfRTRAW}}
\end{figure}

One week of simulated RTRAW data have been generated in DC-1. Each RTRAW file contains about 851 events within 6-second interval. The time interval of each job is determined based on the memory usage and CPU time. One of the challenges encountered in RTRAW production is memory consumption, as the electronics simulation and OEC run together, with multiple events being cached in the memory for event classification. These tasks are executed on the large memory computing nodes. Figure \ref{fig:perfRTRAW} shows the performance of generating RTRAW data. On average, each job consumes 1302 seconds of CPU time and 5.4 GB of memory. The presence of muon shower events will take more memory usage. 

\subsection{Serial reconstruction}
\begin{figure}[h]
    \centering
    \begin{subfigure}{0.48\textwidth}
        \includegraphics[width=\textwidth]{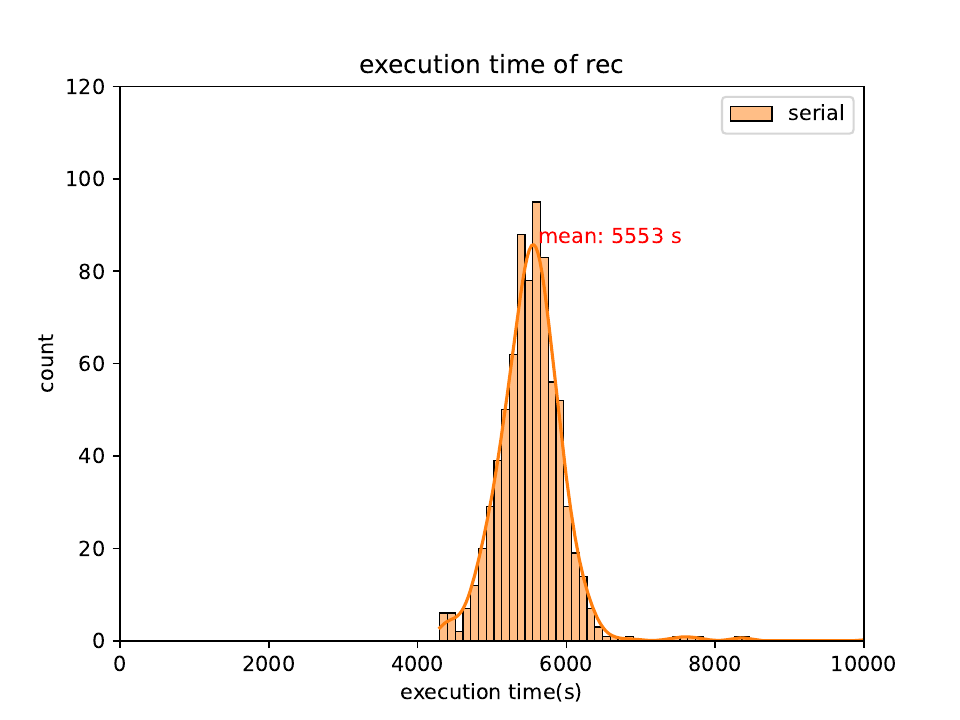}
    \end{subfigure}
    \begin{subfigure}{0.48\textwidth}
        \includegraphics[width=\textwidth]{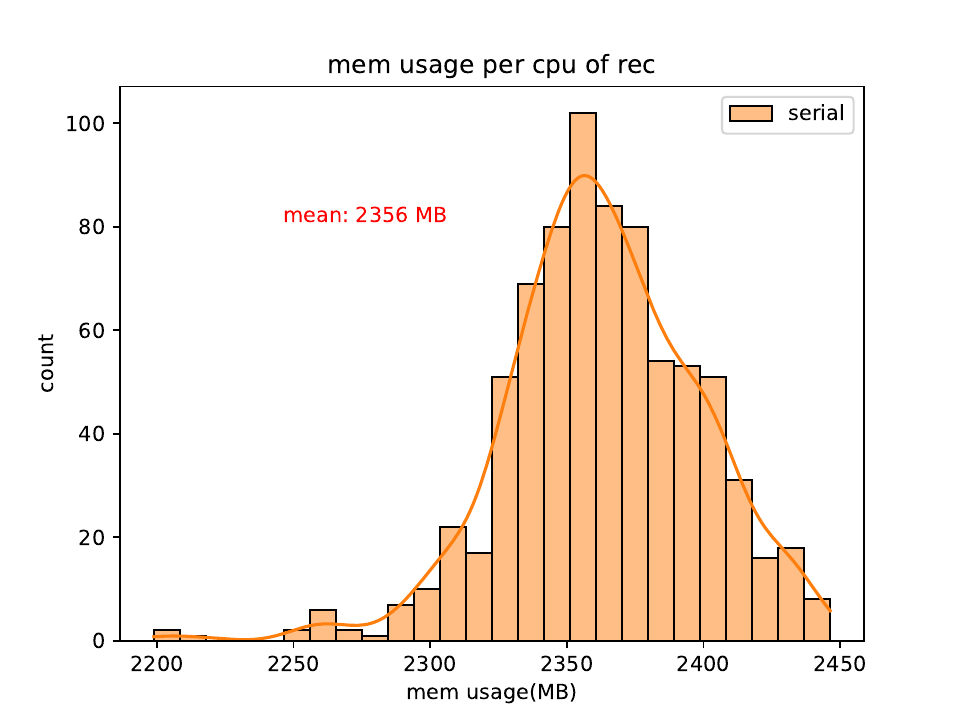}
    \end{subfigure}
    \caption{Performance of serial reconstruction. (a) Execution time. (b) Memory usage. \label{fig:perfREC}}
\end{figure}

The performance of the serial reconstruction is evaluated as a benchmark. As shown in Figure \ref{fig:perfREC}, the average CPU  time required to reconstruct 6-second RTRAW data is 5553 s, with an average memory usage of 2.4 GB. The mean reconstruction speed is about 6.53 seconds per event. The reconstruction of a real data within 80-second interval at 1 kHz will takes about 6 days of CPU time. Consequently, it is necessary to develop multi-threaded reconstruction algorithms to reduce the processing time.

\subsection{Multi-threaded reconstruction}
\begin{figure}[h]
    \centering
    \begin{subfigure}{0.48\textwidth}
        \includegraphics[width=\textwidth]{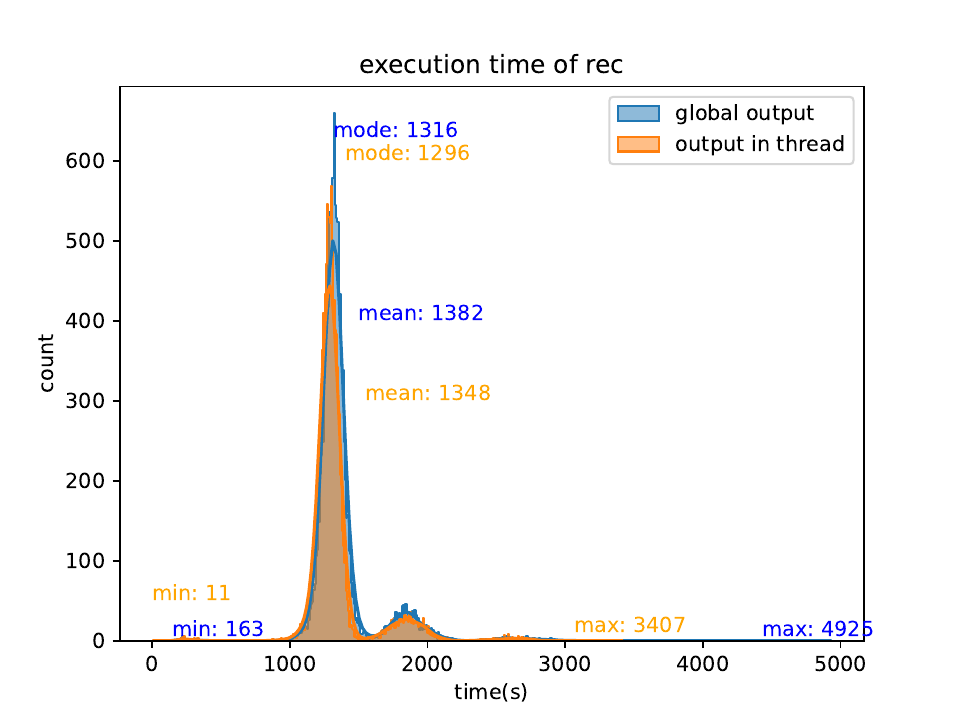}
    \end{subfigure}
    \begin{subfigure}{0.48\textwidth}
        \includegraphics[width=\textwidth]{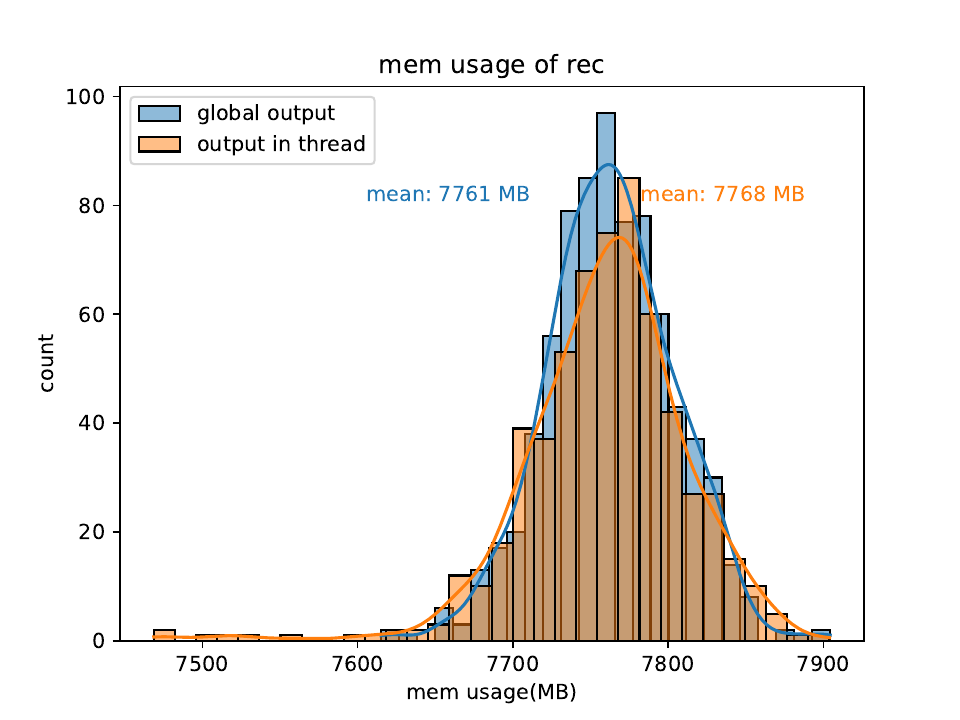}
    \end{subfigure}
    \caption{Performance of multi-threaded reconstruction with 4 CPU cores. (a) Execution time. (b) Memory usage. Note about the second peak in Figure 7 (a): some jobs are scheduled to other computing nodes with different CPU models. \label{fig:perfMTREC}}
\end{figure}

The performance of the multi-threaded reconstruction is evaluated using 4 CPU cores. As shown in Figure \ref{fig:perfMTREC}, the execution time is reduced to a quarter of that required for serial reconstruction. The second peak in the figure is due to the scheduling of jobs on different nodes with varying CPU types within the computing center. The total memory usage is less than 8 GB, which is lower than the total memory usage of 4 different processes.

Given the variability in processing times for different event types and energies, the output could be delayed if an event is not yet completed. Therefore, the jobs are configured with two output modes. In the ``global output'' mode, events are cached in memory in the correct time order, and the data is sequentially saved into a file. In contrast, the ``output in thread'' mode first saves the processed events from different threads into separate files, which are then merged and sorted at the end of the jobs. The figures present the results in these scenarios. The time consumption in ``output in thread'' mode is less than that in the other mode. Therefore, the ``output in thread'' mode is chosen for official data production. 

\section{Conclusions and plans}
The JUNO DC-1 is the first time to test the data processing chain, beginning with RTRAW, mimicking the real data processing. Multi-threaded algorithms have been developed and tested. The database has also been used in both the local cluster and DCI. All primary goals have been successfully met, and ongoing checks on the produced data are in progress. 

However, certain aspects of DC-1 still require enhancement. For instance, only the reconstruction algorithms for the central detector have been tested. These work will be addressed in the upcoming rounds of the JUNO DC. 

\section*{Acknowledgements}
This work is supported by the Strategic Priority Research
Program of the Chinese Academy of Sciences (Grant No. XDA10010900), National Natural Science Foundation of China (12375195), the Youth Innovation Promotion Association, CAS.

\bibliographystyle{iopart-num}
\bibliography{bib}

\end{document}